\documentclass[12pt]{iopart}
\usepackage{graphicx,color}
\usepackage{iopams}
\newcommand{\be}{\begin{equation}} 
\newcommand{\ee}{\end{equation}} 
\newcommand{\bea}{\begin{eqnarray}} 
\newcommand{\eea}{\end{eqnarray}}
\newcommand{\figmf}
{\begin{figure}[htbp]
        \centering
        \includegraphics[angle=0,width=8cm]{match.eps}
		\caption{ Numerical solution shows saturation in 
		Static susceptibility (in units of 10$^4$) vs Temperature curve. This curve
		is in good agreement with Muller's
		experiment in low temperature side, with 
		$\frac{v\delta a^2}{\omega_0^2}= 1, \Delta = 0.003, 
		\eta = 1/\Delta, q_{max} = 0.1$ and at the end $\chi$ and $T$ are rescaled with $0.4/\Delta$  and  $30\Delta$ respectively.The lower curve is the non-selfconsistent fit with the same parameters as the
upper one but with rescalling of $\chi$ and $T$ by 9.5 and 100 respectively.}
	\end{figure}
}
\newcommand{\figQC}
{\begin{figure}[htbp]
       \begin{center}
 \includegraphics[angle=0,width=6.5cm]{delta.eps}
\end{center}
		\caption{ Temperature variation of Susceptibility at different values of $\Delta$ and the log-log plot of the same}
	\end{figure}
}
\newcommand{\figPH}
{\begin{figure}[htbp]
\begin{center}
\includegraphics[angle=0,width=7cm]{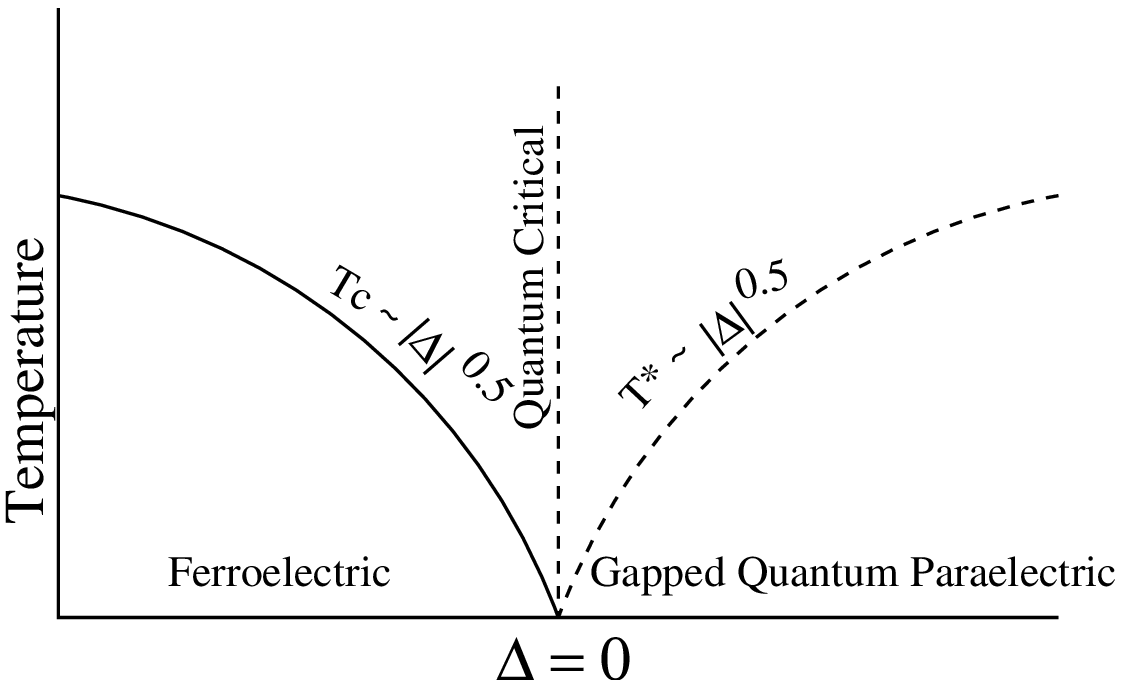}
\end{center}
		\caption{Schematic phase diagram of a typical quantum paraelectric system}
	\end{figure}
}
\begin{document}
\title{Fluctuations and Criticality in Quantum Paraelectrics}
\author{Nabyendu Das$^1$ and Suresh G. Mishra$^{1,2}$}
\ead{nabyendu@iopb.res.in, mishra@iopb.res.in}
\address{$^1$ Institute of Physics, Bhubaneswar 751005 India.\\$^2$National Institute of Science Education and Research, Bhubaneswar, 751005, India}
\begin{abstract}
        The temperature dependence of static dielectric susceptibility
        of a system with strongly coupled fluctuating dipoles 
        is calculated within a self consistent mean fluctuation field approximation. 
        Results are qualitatively in good agreement with, a quantum paraelectric, 
	SrTiO$_3$ in the low temperature regime. We identify this 
        system as a {\it gaped quantum paraelectric} and suggest a  possible experimental 
	realization of a {\it quantum critical paraelectric}  through 
	the application of hydrostatic pressure or doping by impurity.
\end{abstract}
\pacs{77.80.Bh, 64.70.Tg}
\date{\today} 
\submitto{\JPCM}
\maketitle
\section{Introduction and summary}
Phase transitions in displacive systems can not be described by Ising type 
Hamiltonian, which is usually invoked for a system going through order-disorder transition. 
Behavior of these systems are essentially governed by collective oscillations of coupled dipoles 
and the phase transition is described by softening of the corresponding optical mode 
 due to thermal fluctuation ~\cite{Cochran,Anderson}. There are materials like 
SrTiO$_3$, KTaO$_3$ which are supposed to show a displacive transition similar to 
that occurs  BaTiO$_3$, but fail to do so. In fact there is softening of optical 
phonon modes in these materials also, but that does not lead to a transition even 
at very low temperature~\cite{sakudo,muller}. Instead, there is a strong enhancement 
of low temperature dielectric constant. For example, in SrTiO$_3$ the static dielectric 
susceptibility has a very high  saturated value ($\mathcal{O}(10^{4})$) at low temperature 
($ \sim 10K $) followed by a Curie regime ($10$ to  $ 100 $ K) and a long tail 
thereafter. It seems that the classical soft mode concept is insufficient to describe 
various aspects of low temperature behavior. Explanation of the unusual behavior of 
dielectric susceptibility and the stability in the low temperature paraelectric phase has been 
a long standing puzzle. Long ago Barrett 
\cite{barrett} gave a semi-phenomenological theory, which essentially recasts the 
Curie-Weiss formula  with a replacement of temperature $T$ there, by average 
energy, thereby the inverse of dielectric susceptibility could be written as, $\chi^{-1} \propto T_1 \coth({T_1}/{T}) - T_c$, 
where $T_c$ is classically calculated critical temperature and $T_1$ is a quantum scale $\mathcal{O}({\hbar}/{\rm mass})$. This theory, in the high temperature limit, reproduces 
the Curie Weiss law. 
To match experimental data in SrTiO$_3$ the Barretts' formula has been 
found inadequate as one single constant quantum scale $T_1$ can not trace the full curve. The formula has since been modified in various ways, for example, by introducing an extra 
exponent ~\cite{kleeman}, that is, by writing $\chi^{-1}$ as $ (T_1 \coth({T_1}/{T}) - T_c)^{-\nu}$,
and  by making  $T_1$ temperature dependent with an extra scale ~\cite{yuan}, to take care of various
``anomalies'', for example the one near 40K. There has been a proposal 
of attributing this extra energy scale to the structural transition which occurs at 
110K ~\cite{ref-05}. These proposals either follow an order parameter expansion similar to the Landau expansion or some modifications thereof, hence they do not introduce  any new microscopic description. 
In SrTiO$_3$ and  KTaO$_3$ there is no ordering, therefore the low temperature regime where the dielectric constant is enhanced, the physics is dominated by fluctuations of relevant microscopic degrees of 
freedom rather than their averages. In this article we analyze the fluctuation in such systems 
within a  self consistent mean fluctuation field approximation. There are mainly two parameters, 
namely the anharmonicity parameter and the effective stiffness. The zero point or quantum 
fluctuation will be dominant when stiffness is small. The
qualitative behavior of susceptibility is reproduced as well as a new insight gained into the quantum critical behavior of such systems.  A mismatch in theory and experiment curves for the dielectric constant 
at high temperature can be attributed to the effect of structural transition which occurs at higher temperature (i.e. at 110 K in SrTiO$_3$), such discrepancy is irrelevant for the following discussion which refers mainly to the low temperature regime. 

\section{Mean Field Analysis} 
The low temperature physics of SrTiO$_3$ is dominated by fluctuations of Ti ions~\cite{vanderbilt}. The Hamiltonian for such ions is modeled in terms of local quartic oscillators coupled with  a nearest neighbor harmonic interaction ~\cite{Thomas}, 
\begin{equation}
\label{eq:1}
 H= \sum_l \left\lbrace \frac{ p_l ^2 }{2}  +  
	\frac{1}{2}  \omega_0^2  u_l^2  +  \frac{1}{4} \lambda u_l^4 
	\right\rbrace -   \frac{1}{2} \sum_{ll^{\prime}}  
	v  u_l u_{l^{\prime}}.
\end{equation}
The constants $\lambda$ and $v $ are assumed to be positive and mass taken as unity. 
This Hamiltonian describes two local minima with a nearest neighbor coupling  $v$. For 
$|v| <<  |\omega_0^2|$ and $\omega_0^2\ll 0$, it mimics a two state Ising  system with Gaussian fluctuations around one of the local minima. When $|v| \sim   |\omega_0^2|$, there is a possibility of large tunneling between these minima. In this regime the system has to be described in terms of its collective behavior. Such system is called displacive system and the limit $|v| \rightarrow  |\omega_0^2|$ is called Displacive limit. 
In momentum space, 
\be 
         \label{eq:2} 	
             H= \sum_q \frac{1}{2} p_q ^2 + \frac{1}{2} \sum_q ( \omega_0^2 
             - v \delta \cos qa)  u_q u_{-q} + \frac{1}{4} \lambda \sum_{q_1, q_2, q_3} u_{q_1} u_{q_2} u_{q_3}  u_{-q_1 -q_2 - q_3}
\ee
Here $\delta$ is the coordination number and $a$ is lattice spacing.
Now with $
p_q =  \dot{u_q} = - \imath  \omega u_q $ in the kinetic energy term, finally the Hamiltonian within 
the quasi harmonic approximation, i.e.,
\be
\sum_l u_l^4  \approx  6 N (\sigma + \langle u \rangle^2)  \sum_{ q_1 } u_{q_1} u_{- q_1}
\ee
where 
\be
\sigma =  \sum_{q} \langle T u_q (0) u_{-q} (0^+) 
  \rangle,
\ee 
 is  given by,
\be           
H  = \frac{1}{2} \sum_q   (\omega_q^2 - \omega^2) u_q u_{-q}
\ee 
where 
\be
\omega^2_q  =  \omega_0^2 - v \delta \cos qa +  3 \lambda \sigma \simeq  \omega_0^2 - v + v \delta a^2 q^2 + 3 \lambda \sigma
\label{omega_q}
\ee
is the renormalized frequency for small $q$ (such truncation is quite justified for low temperature properties of a near critical system.). We are interested in the paraelectric phase of the system, that is, where $<u> = 0$. Since the system 
is at low temperature and the dielectric constant has an enhanced value,  $<u^2 > $ need not vanish, however. The purpose of present work is to present a self consistent calculation of $<u^2 > $ in 
classical as well as in the quantum regime. The susceptibility, which is related to $<u^2 > $, is 
essentially the phonon propagator,
\be
\chi(q,n)= -\frac{1}{ (\imath \omega_n)^2  - \omega_q^2 },~~
\omega_n= 2n\pi T.
\ee
With $\omega_q$ given by equation(\ref{omega_q}) we have a self consistent equation,
\begin{eqnarray}
  \sigma &=& \sum_{q} \langle T u_q (0) u_{-q} (0^+) 
  \rangle  = \frac{1}{\beta}   \sum_{q,n}  \chi_{qn}    
  e^{ \imath \omega_m 0^{+} } \nonumber \\ 
  &=& \frac{1}{\beta} \sum_{q,n}  
  \frac{1}{ \omega^{2}_n  + \omega_q^2 } \nonumber \\
  &=&\sum_{q} \frac{1}{2\omega_q} \coth  \left(  \frac{\omega_q}{2T} \right)
  \end{eqnarray}
 The solution of this equation will give $\sigma(T)$ which in its asymptotic forms reduces to,
  \bea
 \sigma &=&  \sum_q \frac{T}{\omega_q^2} \sim \int dq~ q^2  \frac{T}{  \omega_0^2 - v 
+ v \delta q^2 + 3 \lambda \sigma }  ~~ {\rm (Classical)} \label{class}\\
  &=&\sum_q \frac{1}{\omega_q} \sim \int dq~ q^2  \frac{1 }{  \sqrt{\omega_0^2  - v + v \delta q^2 + 3 
  \lambda \sigma }} ~~{\rm (Quantum)}. 
\eea
To go into details of temperature dependence of susceptibility, we need to define some physically interesting dimensionless parameters as,  $\Delta = -(\omega_0^2 - v)/\omega_0^2 $,
 $\sigma_c = -(\omega_0^2 - v)/3\lambda$, 
 $\eta = \hbar/(2\omega_0 \sigma_c)$ ($\hbar$ is taken as unity in this article), 
as such $ \Delta \sim \sigma_c \sim \eta^{-1} $.
The parameter $\Delta$ describes the effective stiffness for collective modes at harmonic level. The strength of coupling between various modes near $q =0$ is determined by $\sigma_c^{-1}$   while the parameter $\eta$ tells us about the vicinity to the quantum limit in the system. 
 Introducing normalized 
temperature $x = T/m\omega_0^2\sigma_c$  and using the previously defined parameters, we rearrange the equation (\ref{omega_q}) as follows
\be
\frac{\omega_q^2}{\omega_0^2} =   \frac{v\delta a^2 q^2}{m\omega_0^2} +\Delta(\frac{\sigma}{\sigma_c} -1)
\ee
and
\be
\frac{\sigma}{\sigma_c} = \sum_{q} \frac{\eta\omega_0}{\omega_q} \coth  \left(  \frac{\eta \omega_q}{\omega_0 x} \right).
\label{sigma}
\ee
A self-consistent solution of these equations will give,
\be
\chi(0,0)^{-1} \propto \Delta (\frac{\sigma}{\sigma_c} -1).
\ee
For large enough $\Delta $ the system shows classical behavior, that is $\sigma \sim T$ from equation (\ref{class}). The mode coupling would give corrections higher order in temperature, and $T_c$ 
would be proportional to $\Delta$.
On the other hand as $\Delta$  become smaller and $\eta$ becomes larger and the system move towards the quantum domain. When $\Delta$ or $T_c$ becomes identically zero we have quantum critical point. At this point the zero temperature static long wavelength susceptibility also diverges.
Interestingly the $\Delta =0 $ or $ \omega^2_0 = v$ limit is the {\it displacive limit}, well known in  the structural transition literature. 
A non-self-consistent estimate, with a 
temperature dependent momentum cut-off $(q_{max}\sim T)$, tells that 
$\sigma$ starts from a constant value in the low temperature side and then follows a
 $T^{2}$ behavior in high temperature (up-to Debye temperature) side. Such a non-self-consistent 
estimate gives quite correct result when the system is far away from the quantum critical point i.e. $ |(\omega_0^2 - v)/3\lambda\sigma|<< 1$. At quantum critical point, an estimation of the self-consistent correction is also $\sim T^{2}$. The Barret formula  can not reproduce this result. That formula is essentially outcome of quantum fluctuations of in a single mode theory, which would fail near the quantum critical point as many modes and their coupling would dominate the behavior of system there.This necessitates a self-consistent
 calculation for quantum paraelectrics near its quantum critical point. From figure(1), we learn that the high value of static dielectric susceptibility of SrTiO$_3$ is attributed to the smallness of the parameter
 $\Delta(=0.003)$. This motivates us to treat this system to be near quantum critical point.
The static dielectric susceptibility data of SrTiO$_3$ remind us of  the behavior of itinerant Fermionic systems near quantum phase transition point and fluctuation regime around that. There the (staggered) magnetic susceptibility diverges for (anti-)ferromagnetic transition as the coupling constant crosses a critical value\cite{sreeram}. The case of SrTiO$_3$ is similar to that of liquid Helium-3 \cite{sgm}, where the magnetic susceptibility gets enhanced, as large as ten times, depending upon pressure, from its free Fermionic value. 

\vspace{0.35cm}
\figmf
\section{Quantum Criticality and Hydrostatic Pressure at QCP} 
After the above identification  we now focus on theoretical aspects of quantum criticality in ferroelectric systems.
\figPH
At zero temperature zero mode fluctuation is the most dominant. From the equation (\ref{sigma}) it is clear that zero temperature fluctuation $\sigma_s$ is given by 
\be
 \sigma_s/\sigma_c = \eta \frac{\omega_0}{\omega_q} = \frac{\eta}{\sqrt{\Delta(\sigma_s/\sigma_c -1)}}.
\ee
Writing $\sigma_s/\sigma_c$ as $y$ we get,
\be
y^3 -y^2 - a = 0, ~~~ a = \frac{\eta^2}{\Delta}
\ee
This tells if $ a = 0 $  then $\sigma_s = 0$ is a solution, which is the classical limit. For $a \ne 0$ all solutions
become non-zero and since $y^3 - y^2 = a \ge 0 $ that means $y \ge 1$.
Thus $\sigma_s \ge \sigma_0 $,  moreover the $\sigma_s $ increases as ``$a$'' increases.  The meaning of quantum criticality, in terms of $y$, is $y \rightarrow 1$.  In that limit  the zero temperature properties will show a scaling behavior with (inverse of Correlation Length) $\chi^{-1} \sim \Delta (y - 1) \sim \eta^2\sim(1-v/ \omega_0^2)^{2}$. If we define a \emph{quantum scale} $\xi_Q = (1-v/ \omega_0^2)^{-\frac{1}{2}}$, then $\chi \sim \xi_Q^{-4}$. The point $y = 1$ is essentially the point where effective stiffness ($\Delta$) changes sign. In the regime $\Delta \ge 0 $ self consistency in fluctuation breaks down, system seeks ordering and hence an expansion about the non-zero $<u>$ is required. In this case transition temperature $T_c \sim \Delta^\frac{1}{2}$. On the other hand, in $\Delta \le 0 $ regime the system can not have any ordering and its behavior has to described by self consistent fluctuations. There is a characteristic temperature (cross over temperature in modern parlance \cite{sachdev}) $T^* \sim \Delta^\frac{1}{2}$ which demarcates the boundary between the low temperature {\it gapped quantum paraelectric} behavior and the classical behavior(figure(2)). In case of SrTiO$_3$, the plateau in the susceptibility vs temperature curve is the signature of gapped quantum paraelectric behavior. There is no transition in this system. But there is a crossover from low temperature quantum to high temperature classical behavior at the crossover temperature $T^*$ ($\sim10K$ ). This is exactly the temperature where
plateau ends and the susceptibility curve stars following a Curie behavior(figure(1) and (3)). One can now hope to reach at $\Delta =0$ through tuning some parameters like pressure, impurity etc. The width of this plateau regime vanishes at this point and the system becomes {\it quantum critical}. At this point thermodynamics will be described by power laws in temperature (e. g. $\chi(0,0)^{-1} \sim T^{-2}$) and the system will show some non trivial dynamics. The later is beyond the scope of the present work. It is quite evident here as the controlling factor $v/m\omega_0^2$ strongly depends upon structural aspects and hence this quantum-ness in SrTiO$_3$ can be properly understood through some intrinsic mechanism which give rise to such large tunneling . 

A good possibility of exploring the physics near such quantum critical point is through application
of hydrostatic pressure. Such a technique is already used in case of ferroelectrics and quantum paraelectrics long ago \cite{samara-1} and more recently \cite{samara-2} in different contexts. We found that those experimental results can be discussed more interestingly as is done in the context of itinerant
magnetic system recently\cite{gehring}. Application of hydrostatic pressure will couple to optical mode via its coupling to the acoustic mode. In this case the starting Hamiltonian takes the form
\begin{eqnarray} 
             H&=& \frac{1}{2} \int dq \left[   p_q ^2 +   
	      \left(  \omega_0^2 
             - v \delta \cos qa\right)   u_q u_{-q}\right]  
              + \frac{\lambda}{4}\int \Pi_{i=1}^4 (dq_i u_{q_i}) \delta(\sum_i q_i)~\nonumber\\
&+&g \int dk~dq~ \epsilon(k)~ u_q u_{k-q}  + \frac{K}{2}  \int dq  \epsilon^2 (q) - p\epsilon(0)~
\end{eqnarray}
Here last three terms are results of applications of pressure, in lowest possible order. The parameter ``$g$''  couples strain fields to unit cell displacement related to optic mode and ``$K$'' is the force constant for harmonic acoustic phonons, and the last term shows the coupling of the hydrostatic pressure ``$p$'' to the static strain with some unit strength.
Now if the pressure is strong enough $\epsilon$ has a minima at $\epsilon = \epsilon (0)$ and is given by
\be
\epsilon (0) = p/K.
\ee
Integrating out strain field, we get an effective Hamiltonian  
\bea
 H &=& \int dq \left[  \frac{1}{2} p_q ^2 +   
	     \frac{1}{2} \left(  \omega_0^2 +gp
             - v \delta \cos qa\right)   u_q u_{-q} \right]  \nonumber\\
              &+ & \frac{1}{4} \lambda_R \int \Pi_i dq_i u_{q_1} u_{q_2}
	       u_{q_3}  u_{-q_1 -q_2 - q_3}
\eea
with renormalized coupling constant of quartic term   
\be 
 \lambda_R = \left( \lambda -\frac{2g^2}{K}\right) 
\ee
Again we write a self consistent equation for paraelectric fluctuations as,
\be
\sigma =\int d^dq \frac{1}{\omega_q}\coth \left( \frac{\omega_q}{T}\right) 
\ee
Where
\be	
 \omega^2(q) =  3\Delta \lambda (1+ p/p_0)
              + v \delta q^2a^2/2   
              + 3\lambda_R \sigma {\rm ~and~} p_0 = \frac{3K\Delta\lambda}{g}
\ee
\figQC
Up to this point result is just renormalization of the factor $\Delta$ as $\Delta(1+p/p_0)$ and it becomes an
experimentally controllable parameter. And the behavior of susceptibility at different values of $\Delta$
is shown in the figure.
In this proposal we assume the positivity of $\lambda_R$. Otherwise transition will be
first order and the scaling behavior will not be valid. In real life one can try to induce the effect of negative pressure required in these systems to achieve QCP through some homogeneous effects of non-polar impurity. But in either case nature of the transition can be modified because of strain coupling or disorder respectively. Analysis of such transition in these materials
will be discussed in an upcoming paper\cite{1st}.

\section{Discussion}
We have shown that a mean fluctuation field  theory within a quasi harmonic approximation reproduces the low temperature behavior of susceptibility of a quantum paraelectric quite well. The short range model studied here is justified since only transverse optical modes 
are involved in the ferro-electric fluctuations.  In presence of a long range force longitudinal mode becomes stiff and only transverse modes can get soft. Presence of long range dipolar forces can induce a certain amount of anisotropy in the transverse modes, which can certainly change the critical behavior, however, only with a fairly large value of dipolar contribution to anisotropy in 
the quadratic term \cite{millis}.  We are not aware 
of the first principle results for anisotropy parameters in case of SrTiO$_3$ or KTaO$_3$. 
However, the first principle calculations support our choice of parameter for  the effective stiffness. Compared to BaTiO$_3$ it is about twenty times smaller (Table V in ref \cite{King}) for SrTiO$_3$,
which makes it more near the quantum domain. On the other hand the lattice induced anisotropy in the quartic term is of the same order of magnitude and it would not play a key  role in distinguishing the low temperature behavior in these systems. We leave discussions on anisotropy dependence 
for the future work and stick to isotropic short range model. It is also clear that there is no need to introduce ``anomalous'' regime as proposed earlier. That proposal might be due to the insistence on comparing experimental results with Barrett's formula and its extensions. The experimental behavior is well accounted for in the quantum region and at high temperature the susceptibility smoothly crosses over to the classical behavior. Here we have focused more on physics of low temperature behavior than the exact calculation of various properties. Thus the structural aspects and anisotropy effects are not attempted here. 

\section{Acknowledgment} We thank Professor T.V. Ramakrishnan for introducing us to this subject and for many useful communications. One of us (N.D) would like to thank
Mr. Prabodh Kumar Kuiri for some technical assistance.

\end{document}